\documentclass[aps,prl,reprint,groupedaddress]{revtex4-2}
\usepackage{natbib}
\usepackage{graphicx}
\usepackage{longtable}
\usepackage{MnSymbol}
\usepackage{longtable}
\usepackage{multirow}

\begin{document}
%\preprint{}
\title{Origin of Crystallization Suppression in a New Amorphous Molecular White-Light-Generating Material}
\author{B. D. Klee}
\author{B. Paulus}
\author{J. Link Vasco}
\author{S. Dehnen}
\author{W.-C. Pilgrim}
\email[]{pilgrim@staff.uni-marburg.de}
\affiliation{Department of Chemistry, Philipps-University of Marburg, 35032 Marburg, Germany }
\author{S. Hosokawa}
\affiliation{Department of Physics, Kumamoto University, Kumamoto 860-8555, Japan }
\author{J. R. Stellhorn}
\author{S. Hayakawa}
\affiliation{Department of Applied Chemistry, Hiroshima University, Higashihiroshima 739-8527, Japan}
\date{\today}
\begin{abstract}
The microscopic structure of a new infrared-driven amorphous white-light-generating material was explored by X-ray diffraction, EXAFS and Reverse Monte Carlo simulation. In this material, structural disorder appears to be prerequisite for this nonlinear optical effect. The results are consistent with quantum chemical predictions, but it is also found that the molecular cores are distorted, which is identified as a crystallization inhibitor. Sulfur atoms thereby form a uniform vibrational network, which may be responsible for the high capability of the material to absorb infrared radiation.
\end{abstract}
%\keywords{}
\maketitle
Supercontinuum or White Light Generation (WLG) is one of the most puzzling research areas in nonlinear optics (NLO). The effect is known for long, and defines the spectral broadening of an optic signal propagating through a nonlinear medium. Commercial WLG materials are well investigated and consist of compounds with strong nonlinear optical behavior \cite{alfano_2016, dudley_2006}. They need bulky high-power lasers and are virtually restricted to scientific and medical use. Therefore, the synthesis of a group of much simpler molecular inorganic materials has recently attracted considerable attention, because they produce similar optical response if just irradiated by an unpretentious, inexpensive continuous wave infrared laser diode \cite{Rosemann_2016, Rosemann_2016a, Dehnen_2021}. 
%\label{}
\begin{figure}[b]  
\includegraphics[width=0.9 \linewidth]{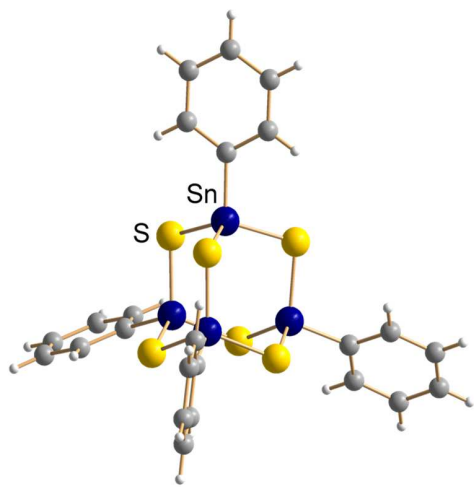}
\caption{\label {figure01} Representation of the [(PhSn)$_4$S$_6$] cluster, the molecular building block of the reportet amorphous WLG.}
%\end{center}
\end{figure}
Such substances are highly desired and could pave the way for a wide range of new and cost-effective applications. They form inversion-symmetry-free heteroadamantane-type clusters of general composition [(RT)$_4$E$_6$], where R is an organic ligand attached to T (Sn, Si, or Ge) which together with E (S or Se) forms the core of the cluster. Figure \ref{figure01} shows the corresponding organotin sulfide cluster [(PhSn)$_4$S$_6$], with phenyl groups attached to the tin atoms.
Two different types of optical response can be generated, depending on the substances’ morphology. When long-range order prevails, and the substances form crystals, they show second-harmonic generation (SHG). If, however, the intermolecular interaction prevents a sufficiently high degree of order, and an amorphous solid is formed, they respond with a brilliant warm-white spectrum to the IR-irradiation, retaining the directionality of the driving laser. No comprehensible explanation has yet been proposed for this, but structural disorder appears to be mandatory for WLG. To understand the elementary mechanisms that determine the morphology in the solid formation and the structural correlations between neighboring molecules, quantum chemical studies have been performed \cite{Hanau_2020, Dehnen_2021} on the interaction of dimers as a minimal model for the amorphous system shown in Figure \ref{figure01} and for the analogous organosilicon sulfide cluster, [(PhSi)$_4$S$_6$], where Sn is replaced by Si, but which forms single crystals and exhibits SHG \cite{Rosemann_2016}. The size of the organic ligands was also varied from phenyl (Ph) to naphthyl (Np) to analyze their influence on the morphology. The studies allowed to decompose the cluster binding energies into core-core, core-ligand, and ligand-ligand interactions. The overall interaction was found to increase with larger ligands while size and shape determine the mutual orientation of the molecules. Larger Np-ligands, each consisting of a flat carbon double-six ring, favor an orientation where they are stacked with respect to each other, while smaller Ph-ligands, as in Figure \ref{figure01}, prefer an alternating, staggered arrangement, where the ligands of one cluster are located in the voids between those of the other cluster. As the most interesting result, it was found that the [(PhSn)$_4$S$_6$] cluster exhibits the by far largest value of core-core interaction. It was also concluded that this interaction is rather isotropic as compared to the [(PhSi)$_4$S$_6$] analogue and responsible for the tendency to form amorphous solids instead of crystals \cite{Dehnen_2021, Hanau_2020}. 
We have therefore started a structure analysis on this amorphous solid to find experimental evidence for the predicted properties and to unravel the origin of the crystallization suppression observed for nearly all Sn-S systems. 
For this we performed X-ray scattering and Extended Absorption Fine Structure (EXAFS) experiments on the amorphous [(PhSn)$_4$S$_6$] material and we have manipulated a Molecular-Reverse-Monte-Carlo Simulation code (m-RMC) to analyze the obtained X-ray and EXAFS data. The RMC code was based on the $RMC\_POT$++ program \cite{Gereben_2012}, which already offers the option to group atoms to a rigid molecule. We have generated a code for the organotin sulfide cluster [(PhSn)$_4$S$_6$] enabling molecular translation and rotation, and free rotation of the phenyl groups \cite{Klee_2018, Klee_2020}.  

Scattering data were measured in transmission at beamline P02.1 \cite{Dippel_2015} of the PETRA III storage ring at DESY, Hamburg using a primary energy of 59.87 keV. The sample was confined in a 1 mm X-ray capillary. Scattered photons were collected using a flat pixel detector and raw data were processed using the software DAWN \cite{Filik_2017}.  Precise corrections due to background-, air-scattering, self-absorption, polarization and Compton contribution were carried out and data were normalized down to $S(Q)$. 
Sn-$K$ edge EXAFS data of the same sample were obtained at beamline P65 \cite{Welter_2019}, also located at PETRA III while S-$K$ edge EXAFS was measured at beamline BL-11 \cite{HAYAKAWA_2008} at the HiSOR facility of the Hiroshima Synchrotron Radiation Center in Japan. Measurements were carried out at room temperature. All scans were performed in transmission mode. Data were analyzed using the Demeter software package (Athena and Artemis) \cite{Ravel_2005}. EXAFS spectra were normalized and background was calculated using the AUTOBK algorithm.

\begin{figure}[ht] 
\includegraphics[width=1.0 \linewidth]{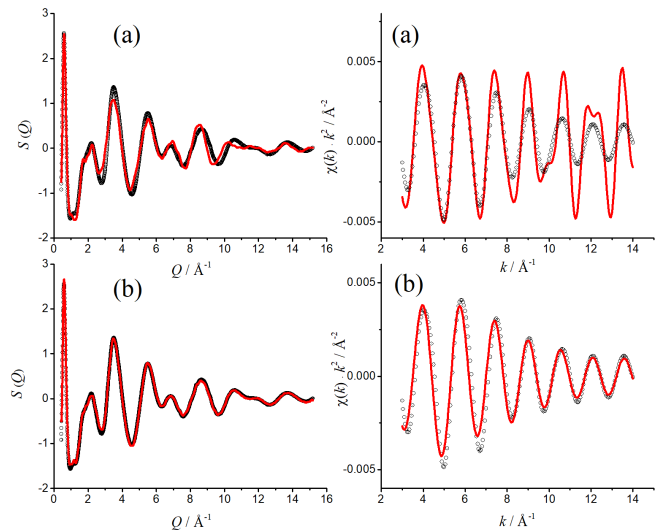}
\caption{\label{figure02} Experimental $S(Q)$ and EXAFS data (black symbols on the left-hand and the right-hand side of a) and b), respectively) as obtained from synchrotron experiments. a) m-RMC results (red lines) from rigid molecules as shown in Figure \ref{figure01}. b) X-ray and EXAFS simulation results (red lines) upon cluster core atoms were additionally allowed to move.}
%\end{center}
\end{figure}

The experimental structure factor $S(Q)$ is depicted on the left-hand side of Figure \ref{figure02} (a) and (b) by the symbols while the experimental Sn-EXAFS, $\chi(k)$ is displayed on the right-hand side also given by symbols. 
In a first RMC simulation, 216 [(PhSn)$_4$S$_6$]-clusters were moved as rigid entities using structural parameters from the DFT-calculated free molecule shown in Figure \ref{figure01} \cite{Rosemann_2016_suppl}.  In addition to the experimental $S(Q)$, the EXAFS was also used as boundary condition. Both functions were iteratively calculated during the simulation and should ideally converge towards the experimental data. Results of this rigid simulation are shown in Figure \ref{figure02} (a) by the red lines. A decent agreement between experimental and simulated $S(Q)$ is already visible. However, there exists a distinctive phase shift between the experimental and simulated $S(Q)$, which becomes readily apparent for $Q$ values above about 8\AA$^{-1}$. It indicates that the molecular structure in the amorphous solid deviates characteristically from the DFT calculated molecule. The difference becomes even more evident when comparing the experimental and simulated EXAFS. Obviously, significant structural differences exist between the single DFT-derived molecule and its counterpart in the amorphous solid. 
In a further simulation, the sulfur and tin atoms of the cluster cores were allowed to move around their DFT-computed positions, while chemically bonded atom pairs were constrained within certain distance ranges. The Sn$\minus$S bond distance was basically left unconstrained (2.05 to 2.65 \AA) because this correlation is prominently represented in the EXAFS and the $S(Q)$ data and comprises hence sufficient information density for the RMC. However, the Sn$\minus$C bond constraint was chosen much tighter (2.05 to 2.25 \AA) to compensate the smaller weighting factor and hence the lower information density. Since the C atoms were not allowed to move, the important consequence was that the Sn atoms kept staying near their initial positions, necessarily keeping the overall molecular structure intact during the simulation. The atomic motions inevitably lead to deviations from the original molecular structure but also to an immediate convergence of both calculated functions towards the experimental curves. They are depicted by the solid lines in Figures \ref{figure02} (b). The agreement with experiment is now excellent in both cases and just small differences between measured and simulated $S(Q)$ remain at lower $Q$ probably resulting from the uncertainty of the exact carbon positions whose weight to the total X-ray scattering is weaker, and the overall structure factor is mainly determined by correlations containing Sn and S.
The difference between (a) and (b) in Figure \ref{figure02} demonstrates that the rigid DFT-calculated molecule is no suitable model to reproduce the experimental data of the amorphous solid. We have therefore explored the simulation box to identify the real-space origin of the observed difference. First, we extracted the average intramolecular atomic distances in the cluster cores from the virtual ensemble. They are compared in Table 1 with spacings obtained from the Sn and S EXAFS experiments. Five different atomic distances exist in the cluster-core, as is visualized in the graphics left hand side of the table. The direct tin-sulfur contact Sn–S$_{short}$ is the only real chemical bond. It can be determined by Sn and S EXAFS, similar to the Sn$\cdots$Sn and S$\cdots$S$_{short}$ distances, each bridging two chemical bonds. However, Sn$\cdots$S$_{long}$ and S$\cdots$S$_{long}$ are respectively second neighbor distances and comprise three bonds. They can no longer be reliably determined by EXAFS. The m-RMC findings are in good agreement with the experimental Sn and S EXAFS values, indicating that the simulated atom positions reflect indeed the structural average in the amorphous solid. Table 1 also shows values for the DFT-calculated cluster. They are slightly larger, indicating a somewhat reduced volume of the molecules’ cluster core in the solid. 

\begin{figure} [ht]
\includegraphics[width=0.9 \linewidth]{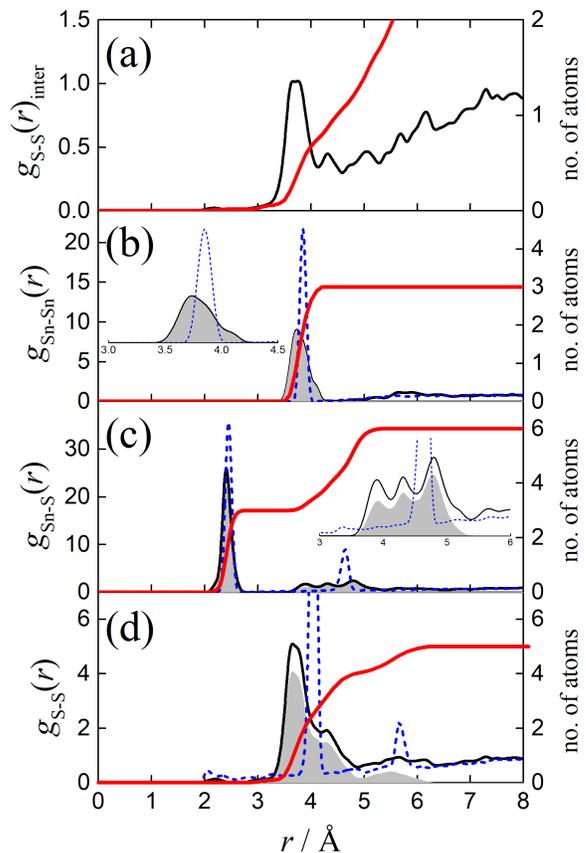}
\caption{\label{figure03} Partial Pair-Distribution-Functions $g_{ij}(r)$ for the cluster-core atoms, as obtained from the rigid m-RMC (blue dashed lines) and the mobile m-RMC (full black line). Shaded areas are intramolecular contribution, full red lines represent number of atoms according to the integral over the intramolecular partial radial distribution function (right scale).}
%\end{center}
\end{figure}

We have also computed the partial pair-distribution functions (PPDF) $g_{ij}(r)$ from the simulation data, where $i$ and $j$ represent either Sn and/or S. They are depicted in Figure 3, as the black solid lines, while the blue dashed curves are PPDFs from the rigid m-RMC, corresponding to the results shown in Figure \ref{figure02}(a). 
\begingroup
\squeezetable
\begin{table}
\caption{\label{table01} Spacings in \AA{} between the atoms of the cluster’s Sn-S core. Spacing designations are displayed in the graphics left to the table.}
\begin{ruledtabular}
\begin{tabular}{ll|cccc}
%\\
%\hspace{5em} & \hspace{2em} & RMC & Sn-EXAFS & S-EXAFS & DFT \\ \hline
\multirow{6} {2.7cm} {\includegraphics[width=2.7cm]{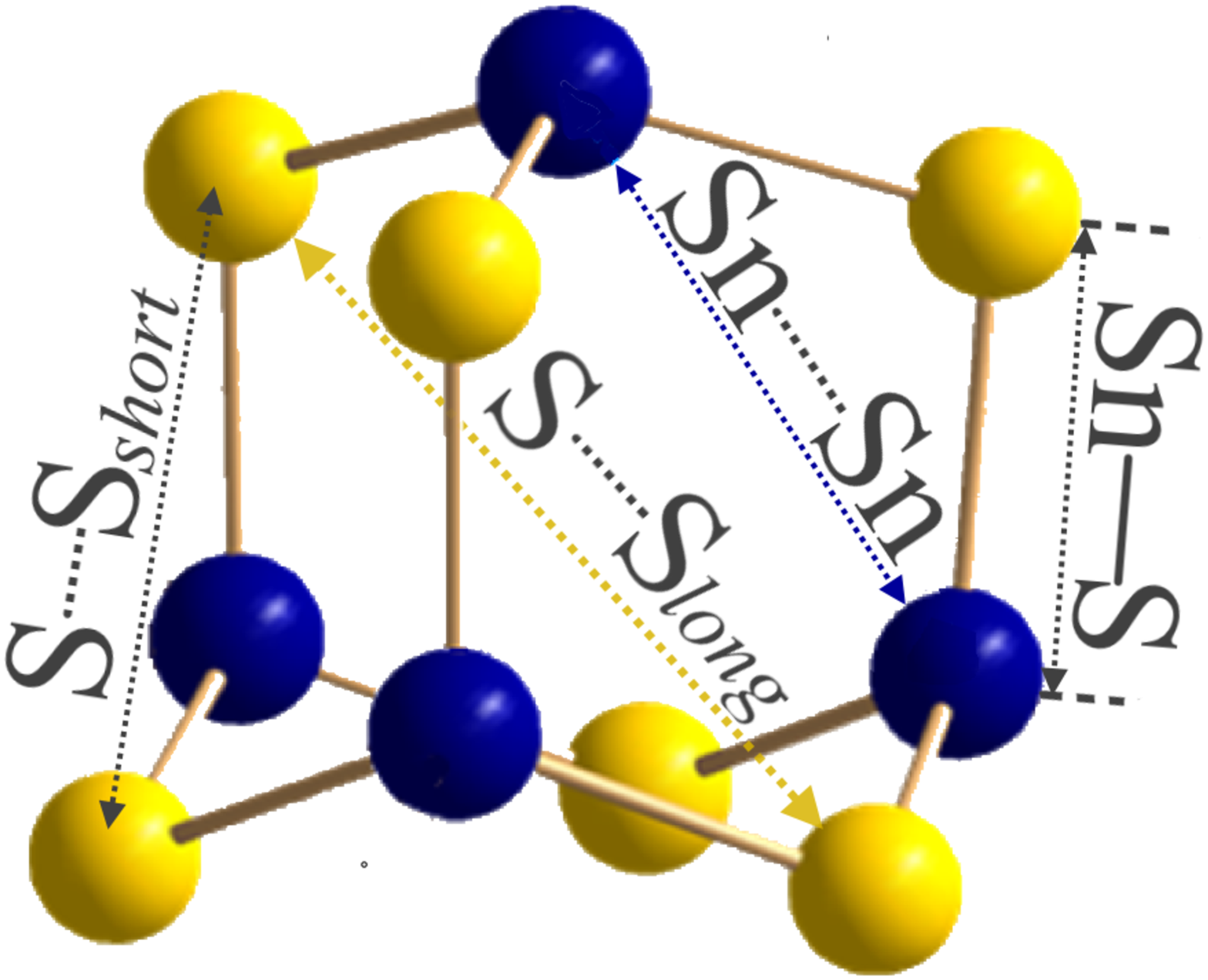}} &  & RMC & Sn-EXAFS & S-EXAFS & DFT \\

\cline{2-6}
 & Sn$\cdots$Sn & 3.8(2) & 3.75(1) & - & 3.84 \\
 & Sn$\minus$S & 2.4(1) & 2.40(1) & 2.39(4) & 2.44 \\
 & Sn$\cdots$S$_{long}$ & 4.5(4) & - & - & 4.63 \\
 & S$\cdots$S$_{short}$ & 3.9(4) & - & 3.9(1) & 4.01 \\
 & S$\cdots$S$_{long}$ & 5.5(3) & - & - & 5.72 \\
%\\
%  &  &  &  &  &  \\
%\hline
\end{tabular}
\end{ruledtabular}
\end{table}
\endgroup
The shaded areas in Figures \ref{figure03} (b)-(d) represent the pure intramolecular contributions and the red solid line (right-hand scale) gives the integral over the partial radial distribution function (PRDF), $4 \pi  r^2 n g(r)  $, with $n$ being the neighboring atom density which defines the number of neighboring atoms hidden under the PPDF peaks. While the average atomic RMC distances in Table \ref{table01} indicate just a slight contraction, they do not provide any evidence for possible deformations of the cluster cores in the amorphous phase or possible correlations to atoms in neighboring molecules.
However, the PDFs show a somewhat different picture: The slight shift of the peak for the Sn$\minus$S bond, from 2.44 to about 2.42 \AA{} is accompanied by a broadening as is seen in Figure \ref{figure03} (c). It is a measure for the uncertainty of the bond length, probably due to thermal vibration. Similarly, is the reduction of the Sn$\cdots$Sn spacing visible in (b) associated with an even stronger broadening. This peak becomes distinctly asymmetric as shown in the enlarged inset of Figure \ref{figure03} (b). It reveals a deformation of the tetrahedral Sn$_4$ frame. The other spacings give further indications of significant distortions: The Sn$\cdots$S$_{long}$ spacing located at 4.63 \AA{}  in the rigid DFT calculated cluster in (c) splits into three components in the amorphous solid which are situated between 3.5 and 5.5 \AA, each containing one single atom. They are also depicted on an enlarged scale in the inset to Figure \ref{figure03}(c). 
Figure \ref{figure03}(d) shows the partial S-S correlations. In the rigid cluster, every sulfur atom is surrounded by four identical neighbors at 4.01 \AA. In the amorphous solid they split into two groups, each containing two atoms, as is indicated by the PRDF integral. One group is represented by a narrower but larger peak at  $\sim$3.68 \AA{} and the second by a smaller but broader peak at  $\sim$4.18 \AA. Each sulfur atom has a further well-defined, more distant single sulfur neighbor at 5.72 \AA{} in the rigid cluster. In the amorphous solid, this correlation is only visible as a very broad peak situated between 5 and 6 \AA, indicating a broad variety of distortions in this specific position. 

The ensemble averages displayed in Figure \ref{figure03}(b-d) clearly show, that the cluster cores in the amorphous [(PhSn$_4$)S$_6$] solid are considerably distorted as compared to its free DFT-computed counterparts, and that it is mainly the positions of the sulfur that have changed. Analyzing the atom-atom correlation between different molecules gives a similarly interesting picture: A pronounced correlation peak centered at $\sim$3.7 \AA{} is found for the intermolecular S-S correlation, depicted in Figure \ref{figure03}(a). It ranges from $\sim$3.2 to $\sim$4.5 \AA{} and the corresponding PRDF-integral reveals one atom, indicating that every cluster core of the RMC ensemble is on average surrounded by one external S-atom on this length scale. It should be noted that such distances compare well with intramolecular S-S spacings (see Table \ref{table01} and Figure \ref{figure03}(d)).
 
\begin{figure} [ht]
\includegraphics[width=0.9 \linewidth]{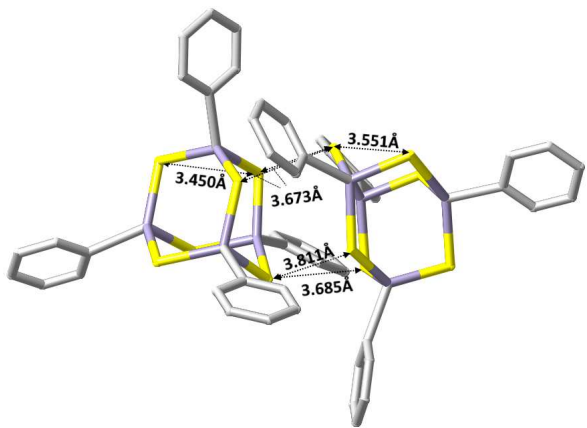}
\caption{\label{figure04} Arbitrarily selected dimer of the simulation box. Yellow:  positions of S atoms. Purple: positions of Sn atoms. Nearest inter- and intramolecular S-S spacings are denoted.}
%\end{center}
\end{figure}

Coordinates of the molecular centers of mass were determined for the [(PhSn$_4$)S$_6$] cluster cores from the simulation box and a next neighbor average distance between such centers could be determined to give 7.5(1) \AA. About 25\% of all clusters are bound in dimers within this spacing. Higher molecular chains, partly branched, also exist. The dimer distance is larger than theoretically predicted ($6-6.5$ \AA{}) \cite{Hanau_2020} because in the dense solid the interaction is shared by more than one neighbor. The spacing is also somewhat larger than in the [(PhSi)$_4$S$_6$] crystal, where the slightly smaller Si-S cluster cores are also arranged in dimers with spacings of 7.05 to 7.46 \AA{} \cite{Hanau_2020}. The mutual arrangement of the phenyl ligands in our [(PhSn)$_4$S$_6$] dimers is an alternating staggered configuration, as predicted by theory and which is also found in crystalline [(PhSi)$_4$S$_6$] \cite{Hanau_2020}. Figure \ref{figure04} shows one example of a [(PhSn)$_4$S$_6$] dimer arbitrarily chosen from our simulation box. 
However, despite some similarity with the arrangement in the [(PhSi)$_4$S$_6$] crystal, the two systems differ significantly. In the crystal, the clusters are undistorted Sn-S adamantanes and intra molecular S$\cdots$S$_{short}$ spacings (3.55 \AA) and intermolecular S$\cdots$S distances (3.84 \AA) can clearly be distinguished. In the amorphous solid the [(PhSn)$_4$S$_6$] clusters are distorted, and intermolecular S$\cdots$S spacings as short as those for intramolecular S$\cdots$S$_{short}$ distances are found in every dimer of the simulation box. Such values are also displayed in the example of Figure \ref{figure04}. Apparently, more S$\cdots$S$_{short}$ spacings are needed to reproduce the X-ray and EXAFS pattern than are present inside the clusters. Therefore, inter-cluster S$\cdots$S spacings are employed, which is only possible by appropriate deformation since the organic ligands prevent a closer approach between the cluster cores. In the real solid, this process is driven by the strong core-core interactions  identified previously \cite{Hanau_2020}. This also explains the high degree of isotropic interaction between the cluster cores reported in \cite{Hanau_2020}, which was made responsibel for the tendency to suppress crystallization. 
The degree of deformation varies from cluster to cluster, and so do the force fields acting between them. The whole cluster ensemble behaves like a distribution of slightly different molecules for which it is difficult to find a common state of crystalline order with minimum free energy. The various distortions therefore act as a crystallization suppression. In fact, nearly all clusters with {Sn$_4$S$_6$} cores that were synthesized so far form only amorphous solids \cite{Dehnen_2021}. Only one crystalline example exists, namely with benzyl groups as ligands. However, its crystal structure shows no dimer formation and is determined from interactions between the organic ligands, which are arranged regularly in the crystal lattice \cite{Dornsiepen_2019}.  
The sulfur atoms in the amorphous [(PhSn)$_4$S$_6$] solid try to distribute themselves uniformly like a close-meshed net, and it is tempting to identify this as a vibrational network. This would offer the possibility of generating corresponding phonons over a wider range of $k$ values, and could thus explain the high receptivity of the material to the IR laser field.

To conclude, we have explored the structure of a new amorphous solid exhibiting white-light-generation, and we found good agreement with quantum chemical predictions. However, in contrast to the DFT-based calculations, distinct distortions of the molecular cluster cores are observed, and identified as the origin for the isotropic inter-molecular interaction, resulting from different distortions and orientations of the clusters. These distortions are responsible for the tendency of molecules with {Sn$_4$S$_6$} cores to suppress crystal formation. Thereby, sulfur atoms play a special role as the difference between  intra- and intermolecular distances vanishes. Instead of constant spacing, they strive for a spatial distribution resembling a vibrational network, which could  explain the high receptivity of the material for IR radiation.
\\
\\
\begin{acknowledgments}
The authors acknowledge Dr. Michael Wharmby for assistance and support at the X-ray scattering experiments on P02.1. We also express our gratitude for funding by the German Research Foundation (Deutsche Forschungsgemeinschaft, DFG), Grant No. 398143140, related to the Research Unit FOR 2824. We also acknowledge the great working conditions and support of the following large-scale facilities: German Electron-Synchrotron (Deutsche Elektronen-Synchrotron, DESY, a member of the Helmholtz Association HGF), beamlines P65 (proposal ID I-20190122), P02.1 (rapid access program 2021A, proposal ID RAt-20010143) and the HiSOR facility of the Hiroshima Synchrotron Radiation Center (BL-11, proposal no. 20AG034). 
\end{acknowledgments}


\begin{thebibliography}{16}

\bibitem{alfano_2016}
R.~R. Alfano, The Supercontinuum Laser Source, Springer, New York, 2016.
%\newblock \href {https://doi.org/10.1007/978-1-4939-3326-6}
% {\path{doi:10.1007/978-1-4939-3326-6}}.

\bibitem{dudley_2006}
J.~M. Dudley, G.~Genty, S.~Coen, Reviews of Modern Physics 78 (2006) 1135.
%\newblock \href {https://doi.org/10.1103/revmodphys.78.1135}
%{\path{doi:10.1103/revmodphys.78.1135}}.

\bibitem{Rosemann_2016}
N.~W. Rosemann, J.~P. Eu{\ss}ner, A.~Beyer, S.~W. Koch, K.~Volz, S.~Dehnen, 
S.~Chatterjee, Science 352 (2016) 1301.
%\newblock \href {https://doi.org/10.1126/science.aaf6138}
%{\path{doi:10.1126/science.aaf6138}}.

\bibitem{Rosemann_2016a}
N.~W. Rosemann, J.~P. Eu{\ss}ner, E.~Dornsiepen, S.~Chatterjee, S.~Dehnen, 
Journal of the American Chemical Society 138 (2016) 16224.
%\newblock \href {https://doi.org/10.1021/jacs.6b10738}
%{\path{doi:10.1021/jacs.6b10738}}.

\bibitem{Dehnen_2021}
S.~Dehnen, P.~R. Schreiner, S.~Chatterjee, K.~Volz, N.~W. Rosemann, W.-C. Pilgrim, D.~Mollenhauer, S.~Sanna,
ChemPhotoChem 5 (2021) 1029.
%\newblock \href {https://doi.org/10.1002/cptc.202100260}
%{\path{doi:10.1002/cptc.202100260}}.

\bibitem{Hanau_2020}
K.~Hanau, S.~Schwan, M.~R. Schäfer, M.~J. Müller, C.~Dues, N.~Rinn, S.~Sanna, S.~Chatterjee, D.~Mollenhauer, S.~Dehnen, 
%Towards understanding the reactivity and optical properties of organosilicon sulfide clusters,
Angewandte Chemie 133 (2020) 1196.
%\newblock \href {https://doi.org/10.1002/ange.202011370}
%{\path{doi:10.1002/ange.202011370}}.

\bibitem{Gereben_2012}
O.~Gereben, L.~Pusztai,
%{RMC}{\_}{POT}: A computer code for reverse monte carlo modeling the structure of disordered systems containing molecules of arbitrary complexity, 
Journal of Computational Chemistry 33 (2012) 2285.
%\newblock \href {https://doi.org/10.1002/jcc.23058}
%{\path{doi:10.1002/jcc.23058}}.

\bibitem{Klee_2018}
B.~D. Klee, E.~Dornsiepen, J.~R. Stellhorn, B.~Paulus, S.~Hosokawa, S.~Dehnen, W.-C. Pilgrim,
%Structure determination of a new molecular white-light source
%(phys. status solidi b 11/2018), 
physica status solidi (b) 255 (2018) 1870138.
%\newblock \href {https://doi.org/10.1002/pssb.201870138}
%{\path{doi:10.1002/pssb.201870138}}.

\bibitem{Klee_2020}
B.~D. Klee, B.~Paulus, S.~Hosokawa, M.~T. Wharmby, E.~Dornsiepen, S.~Dehnen, W.-C. Pilgrim,
%Generating large starting configurations for molecular reverse monte carlo modelling of an unique non-linear optical amorphous solid,
Journal of Physics Communications 4 (2020) 035004.
%\newblock \href {https://doi.org/10.1088/2399-6528/ab756c}
%{\path{doi:10.1088/2399-6528/ab756c}}.

\bibitem{Dippel_2015}
A.-C. Dippel, H.-P. Liermann, J.~T. Delitz, P.~Walter, H.~Schulte-Schrepping, O.~H. Seeck, H.~Franz,
%Beamline p02.1 at {PETRA} {III} for high-resolution and high-energy powder diffraction, 
Journal of Synchrotron Radiation 22 (2015) 675.
%\newblock \href {https://doi.org/10.1107/s1600577515002222}
%{\path{doi:10.1107/s1600577515002222}}.

\bibitem{Filik_2017}
J.~Filik, A.~W. Ashton, P.~C.~Y. Chang, P.~A. Chater, S.~J. Day, M.~Drakopoulos, M.~W. Gerring, M.~L. Hart, O.~V. Magdysyuk, S.~Michalik, A.~Smith, C.~C. Tang, N.~J. Terrill, M.~T. Wharmby, H.~Wilhelm,
%Processing two-dimensional x-ray diffraction and small-angle scattering data {inDAWN} 2,
Journal of Applied Crystallography 50 (2017) 959.
%\newblock \href {https://doi.org/10.1107/s1600576717004708}
%{\path{doi:10.1107/s1600576717004708}}.

\bibitem{Welter_2019}
E.~Welter, R.~Chernikov, M.~Herrmann, R.~Nemausat,
%A beamline for bulk sample x-ray absorption spectroscopy at the high brilliance storage ring {PETRA} {III}, 
AIP Conference Proceedings, 2054 (2019) 040002.
%\newblock \href {https://doi.org/10.1063/1.5084603}
%{\path{doi:10.1063/1.5084603}}.

\bibitem{HAYAKAWA_2008}
S.~Hayakawa, Y.~Hajima, S.~Qiao, H.~Namatame, T.~Hirokawa,
%Characterization of calcium carbonate polymorphs with ca k edge x-ray absorption fine structure spectroscopy,
Analytical Sciences 24 (2008) 835.
%\newblock \href {https://doi.org/10.2116/analsci.24.835}
%{\path{doi:10.2116/analsci.24.835}}.

\bibitem{Ravel_2005}
B.~Ravel, M.~Newville,
%{ATHENA},{ARTEMIS},{HEPHAESTUS}: data analysis for x-ray absorption spectroscopy {usingIFEFFIT}, 
Journal of Synchrotron Radiation 12 (2005) 537.
%\newblock \href {https://doi.org/10.1107/s0909049505012719}
%{\path{doi:10.1107/s0909049505012719}}.

\bibitem{Rosemann_2016_suppl}
N.~W. Rosemann, J.~P. Eu{\ss}ner, E.~Dornsiepen, S.~Chatterjee, S.~Dehnen,
%Organotetrel chalcogenide clusters: Between strong second-harmonic and white-light continuum generation,
Journal of the American Chemical Society 138 (2016) 16224 (supp. Information).
%\newblock \href {https://doi.org/10.1021/jacs.6b10738}
%{\path{doi:10.1021/jacs.6b10738}}.

\bibitem{Dornsiepen_2019}
E.~Dornsiepen, F.~Dobener, S.~Chatterjee, S.~Dehnen, 
%Controlling the white-light generation of [({RSn}) 4 e 6 ]: Effects of substituent and chalcogenide variation, 
Angewandte Chemie 131 (2019) 17197.
%\newblock \href {https://doi.org/10.1002/ange.201909981}
%{\path{doi:10.1002/ange.201909981}}.

\end{thebibliography}
\end{document}